\documentclass[12pt]{article}
\usepackage[dvips]{graphicx}
\usepackage{pdproc}

  \makeatletter 
  \def\@cite#1{[#1]} 
  \makeatother    
   \def\gsim{\lower0.5ex\hbox{$\:\buildrel >\over\sim\:$}}
   \def\lsim{\lower0.5ex\hbox{$\:\buildrel <\over\sim\:$}}
   
   \def \rp{R_P}
    \def \lp{{L\!\!\!/}}
\begin{document}

\renewcommand{\thefootnote}{\alph{footnote}}

\title{
Radiative generation of R-parity violating Yukawa-like interactions
from supersymmetry breaking
}

\author{ Sourov Roy}

\address{ 
Helsinki Institute of Physics, University of Helsinki \\
P.O. Box 64 (Gustaf H\"allstr\"omin Katu 2)¶, Helsinki, FIN-00014, 
Finland
\\ {\rm E-mail: roy@pcu.helsinki.fi}}

\abstract{
Yukawa-like R-parity violating (RPV) couplings are generated  
as effective operators with a typical strength of ${\cal O}(10^{-3})$ 
through loops involving the scalars, the gauginos and the soft 
supersymmetry breaking RPV interactions. Neutrino masses and other 
phenomenological implications of such a scenario are discussed. 
}

\normalsize\baselineskip=15pt

\section{Introduction}

In general, the SUSY Lagrangian can have Yukawa-like 
trilinear R-parity violating interactions \cite{RPVreview} with 
a strength of ${\cal O}(1)$. However, strong constraints have been 
derived on these couplings from various experimental searches which 
suggest that as long as the masses of the superpartners are in the range 
of a TeV or less then these RPV couplings should be much smaller than 
${\cal O}(1)$. In this talk I describe a scenario of generating RPV 
violating Yukawa-like effective operators, whose structures are similar 
to the ones conventionally parametrized by $\lambda,~\lambda^\prime$ and
$\lambda^{\prime \prime}$ in the superpotential but with a strength 
typically much less than ${\cal O}(1)$ \cite{Shaouly-sou}. We start 
with the assumption that the superpotential conserves R-parity and 
that RPV is introduced in the SUSY Lagrangian through the ``soft" trilinear
scalar operators which break supersymmetry. These soft scalar trilinear 
RPV operators cannot renormalize the Yukawa-like RPV operators and 
keep the latter zero at all scales. However, such a non-vanishing 
low-energy RPV soft operator can generate an effective Yukawa-like 
RPV interaction at the one-loop level involving the soft interaction 
and the gauginos and the scalars which run in the loop. This 
way one expects that in general these effective RPV interactions will 
show additional patterns (compared to the $\lambda,~\lambda^\prime$ 
and $\lambda^{\prime \prime}$ ones) due to their explicit dependence 
on the particles running in the loops.

\section{R-parity violating effective operators}

Let us assume for the purpose of illustration that R-parity is violated 
in the Lagrangian {\it only} through the pure leptonic (lepton number 
violating) soft SUSY breaking operators:

\begin{eqnarray}
\Delta V_{\rp,\lp}^{soft} = \epsilon_{ab} \frac{1}{2}
a_{ijk} {\tilde L}^a_i {\tilde L}^b_j {\tilde E}_k^c + {\rm h.c.}~,
\label{slp}
\end{eqnarray}

\noindent where
${\tilde L}({\tilde E}^c)$ are the scalar components
of the leptonic SU(2)
doublet(charged singlet) supermultiplets ${\hat L}({\hat E}^c)$,
respectively,
$\tilde L = (\tilde\nu_L, \tilde e_L)$ and
${\tilde E}^c= \tilde e_R$. Also,
$a_{ijk}=-a_{jik}$,
due to the SU(2) indices $a,~b$.

This means that in our framework the superpotential conserves
$\rp$. For example, there is no such term like $\epsilon_{ab}
\lambda_{ijk} {\hat L}^a_i {\hat L}^b_j {\hat E}_k^c/2$ in the
superpotential. 
 
\begin{figure}[htb]
\begin{center}
\includegraphics*[width=4cm]{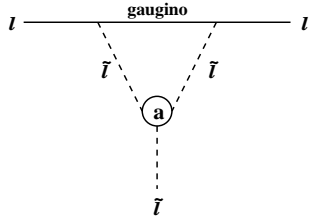}
\caption{%
A typical one-loop diagram generating
an effective $\lambda$-like operator.
$\tilde\ell$ is either a charged slepton
or a sneutrino and $\ell$ is a lepton or a neutrino. $a$ is
the trilinear soft breaking coupling defined in (\ref{slp}).}
\label{fig1}
\end{center}
\end{figure}

We define the effective RPV terms which are generated
at one-loop through the soft operator in (\ref{slp})
via diagrams of the type shown in Fig.~\ref{fig1} as follows:

\begin{eqnarray}
{\cal L}_{\rp}^{eff} Â&\equiv&
\frac{1}{2} \left( \frac{a_{ijk}}{16 \pi^2} \right) [
\tilde\nu_{Li} {\bar e}_k \left( A_{L,ijk} L + A_{R,ijk} R \right) e_j
+  B_{L,ijk} \tilde\nu_{Li} {\bar\nu}_k L \nu_j  
\nonumber \\[4mm]& & \hfill
+C_{L,ijk} {\tilde e}_{Lj} {\bar e}_k L \nu_i
+ D_{L,ijk} {\tilde e}_{Rk} {\bar\nu}_i^c  L  e_j
- (i \to j) ] +{\rm h.c.},
\label{effrpv}
\end{eqnarray}
\noindent where $i,j,k$ are generation indices and
$L(R) \equiv (1-(+)\gamma_5)/2$ are the chirality projection operators.

We find that $B_{L,ijk} \propto m_\nu$, i.e.,
no ${\tilde\nu} \bar\nu \nu$ term in the limit of zero neutrino
masses. Also, $A_{R,ijk} \ll A_{L,ijk}$ since $A_{R,ijk}$ is 
proportional to the leptonic Yukawa couplings and we neglect them. 
The expressions for the remaining form factors $A_{L,ijk}$, $C_{L,ijk}$ 
and $D_{L,ijk}$ can be found in Ref. \cite{Shaouly-sou}. 
 
In order to realize this scenario, let us suppose that SUSY
breaking occurs spontaneously in a hidden sector at the scale
$\Lambda \sim 10^{10}-10^{11}$ GeV, described by the $\rp$-conserving
superpotential:
      
\begin{eqnarray}
W=m_{12}\hat\Phi_1\hat\Phi_2+g\hat\Phi_3\left( \hat\Phi_2^2-M^2 \right)
\label{FOM}~,
\end{eqnarray}

\noindent where $m_{12} \sim M \sim \Lambda$ and under $\rp$,
$\hat\Phi_1,\hat\Phi_2,\hat\Phi_3 \to
-\hat\Phi_1,-\hat\Phi_2,\hat\Phi_3$.
SUSY breaking is triggered by
the vacuum expectation values (VEV's) of the
auxiliary F-term ($F_{\Phi_i}$) of $\hat\Phi_1,\hat\Phi_2,\hat\Phi_3$.
Supergravity mediation of SUSY breaking can then be parametrized
by the following $\rp$-conserving superpotential \cite{Shaouly-sou}:

\begin{eqnarray}
{\frac 1 {M_{Pl}}} \int d^2 \theta \left[ {\hat \Phi_1}{\hat L}{\hat
L}{\hat
E^c} + {\hat \Phi_2}{\hat L}{\hat L}{\hat E^c} \right]
+ h.c.~, 
\label{sup}
\end{eqnarray}
 
\noindent which will spontaneously break $\rp$ and induce
the soft operator in (\ref{slp}) with
$a \sim m_W$. The superpotential in (\ref{sup})
will also generate the operators
$\propto \lambda$ with an extremely suppressed
coupling: $\lambda \sim 10^{-9}-10^{-8}$ at the high scale,
essentially causing the soft operator in (\ref{slp}) to be
the only source for RPV in this model.

One should also note that the lepton number violating soft bilinear term
$B_i \tilde L_i H_u$ will be radiatively (one-loop) generated 
\cite{Shaouly-sou} by the non-zero soft trilinear $a$ term in (\ref{slp}). 
In the leading log and in the limit of vanishing $\lambda$ and $\mu_i$ 
(bilinear RPV terms in the superpotential), 
 
\begin{equation}
B_i(M_Z) \sim -\frac{1}{16\pi^2} h_\tau(M_Z) \mu(M_Z) a_{i33}(M_Z)
\ln \left(\frac{M_{Pl}^2}{M_Z^2}\right) \,
\label{bi}
\end{equation}
 
\noindent where $h_\tau$ is the $\tau$ Yukawa coupling and $\mu$
is the Higgsino mass parameter ($\mu \hat H_u \hat H_d$). R-parity
violation only by soft terms has also been discussed recently in 
Ref. \cite{jones}. 

\section{Some numerical results}

\begin{table}[h]
\begin{center}
\caption{
Values for the effective
RPV form factors $A_{L,ijk}$, $C_{L,ijk}$ and $D_{L,ijk}$
within the Snowmass 2001 benchmark points SPS1, SPS2, SPS4 and 
SPS5 of the mSUGRA parameter space. 
Taken from Ref.~\protect\cite{Shaouly-sou}.
}
\begin{tabular}{|c|c|c|c|c|}
\hline
Effective form factor &SPS1&SPS2&SPS4&SPS5\\
(${\rm GeV}^{-1}$) & & & & \\
\hline
$|A_{L,ijk}| \times 10^4 $ & $3.5-3.6$ & $0.08$  & $0.8-1.3$  &
$2.5-2.6$ \\
$|C_{L,ijk}| \times 10^4 $ & $3.4-3.5$ & $0.08$   & $0.7-1.1$ &
$2.5-2.6$    \\
$|D_{L,ijk}| \times 10^4 $ & $6.8$    & $0.3$  & $2.0-2.6$ & $5.2-5.3$
\\
\hline
\end{tabular}
\label{tab1} 
\end{center}
\end{table}

\noindent In Table \ref{tab1} we give a sample of our numerical results
for the three effective RPV form factors, corresponding to
the ``Snowmass 2001'' benchmark points SPS1, SPS2, SPS4 and SPS5 of the
mSUGRA scenario \cite{Shaouly-sou}. We see that the SPS1 and SPS5 scenarios
give the largest effective couplings, of the order of $10^{-4}-10^{-3}$
if $a_{ijk} \sim 16 \pi^2$ GeV $\sim 150$ GeV.
The existing upper bounds \cite{RPVreview} on $\lambda_{ijk}$ 
are larger than the expected values of our effective couplings 
if $a_{ijk}$ is of the order of the electroweak mass scale. 

\begin{figure}[htb]
\begin{center}
\includegraphics*[width=7cm]{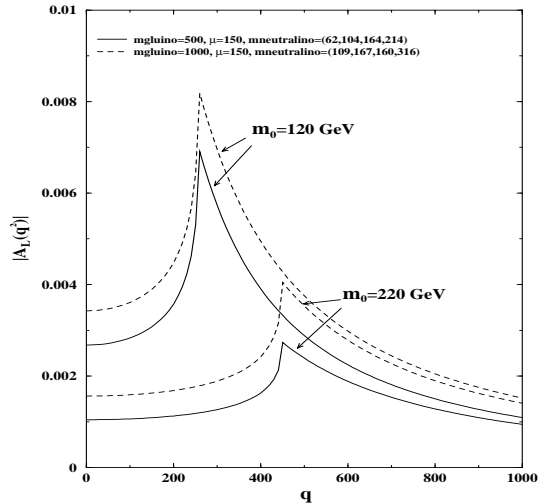}
\caption{%
Momentum dependence of the form factor $|A_L(q^2)|$ for two different
values of the scalar masses ($m_0$) in the loop. All masses are in GeV.
}
\label{fig2}
\end{center}
\end{figure}

One should also note that the form factors are momentum dependent. In 
Fig.\ref{fig2}, we have shown the momentum (q) dependence of the form 
factor $|A_L(q^2)|$. From this figure we see a large enhancement at a 
value of q where the sleptons inside the loop can be produced on-shell. 
This is one of the very interesting features of this scenario and could 
possibly be tested, for example, by the process $e^+e^- \rightarrow 
\tau^+ \tau^-$ through an s-channel ${\tilde \nu}_\mu$ exchange. If such 
an enhancement is observed in $e^+e^- \rightarrow \tau^+ \tau^-$, then 
one should also be able to detect the on-shell production of a pair of 
sleptons that run in the loop, i.e., $e^+e^- \rightarrow 
{\tilde l}~{\tilde l}$.  

\noindent Another interesting way to test this scenario is to look at
the ratio of the two partial decay widths $\Gamma({\tilde \nu}_\tau
\rightarrow \mu^+ \mu^-)$ and $\Gamma({\tilde \mu^-}_L \rightarrow {\bar
\nu}_\tau \mu^-)$. In our scenario this ratio is given by (neglecting
the lepton mass)

\begin{equation}
{{\Gamma({\tilde \nu}_\tau \rightarrow \mu^+ \mu^-)} \over
{\Gamma({\tilde \mu^-}_L \rightarrow {\bar \nu}_\tau \mu^-)}}
= \left(m_{{\tilde \nu}_\tau}/m_{{\tilde \mu}_L}\right)
{{|A_{L,322}|^2} \over {|C_{L,322}|^2}}, \,
\end{equation}       
\noindent whereas for the conventional tree level RPV $\lambda$-type 
coupling this ratio is given by $m_{{\tilde \nu}_\tau}/m_{{\tilde \mu}_L}$. 
The measurement of this ratio of the two partial decay widths as a function 
of the mass ratio $m_{{\tilde \nu}_\tau}/m_{{\tilde \mu}_L}$ can help us 
understand the underlying scenario.

Let us now briefly discuss the generation of neutrino mass in our model.
The soft RPV $a$-term in (\ref{slp}) gives rise to a neutrino mass only 
at the two-loop level (see \cite{burzumaty}), or equivalently, as an 
effective ``tree-level" neutrino mass which is generated by the sneutrino 
VEVs ($v_i$) and is $\propto v^2_i$. Since $v_i \propto B_i/M_{SUSY}$ 
\cite{hirsch} and in our model $B_i$ is a one-loop quantity, clearly 
this also is essentially a two-loop effect. Similarly, ``one-loop" neutrino 
masses involving two vertices of our effective $\lambda$s are infact
coming from three-loop diagrams.

\section{Acknowledgements}

I thank Shaouly Bar-Shalom for numerous helpful discussions and 
collaboration in Ref.\cite{Shaouly-sou} on which this talk was based. 
I would also like to thank the organizers of SUSY04 for the hospitality.
This work is supported by the Academy of Finland (Project numbers 104368 
and 54023).   

\bibliographystyle{plain}

\end{document}